\documentclass[twocolumn,aps,prl,groupedaddress]{revtex4}
\usepackage{epsfig,amssymb,amsmath}
\usepackage[hypertex,linkcolor=red]{hyperref}
\def\comment#1{}
\begin{document}
\title{A quantum solution to the arrow-of-time dilemma: Reply}
\author{Lorenzo Maccone}\affiliation{MIT, 32 Vassar Street, Cambridge,
  MA 02139}

\begin{abstract}
  I acknowledge a flaw in the paper ``A quantum solution to the arrow
  of time dilemma'': as pointed out by Jennings and Rudolph,
  (classical) mutual information is not an appropriate measure of
  information.  This can be traced back to the quantum description
  underlying my analysis, where quantum mutual information is the
  appropriate measure of information. The core argument of my paper
  (summarized in its abstract) is not affected by this flaw.
  Nonetheless, I point out that such argument may not be adequate to
  account for all phenomena: it seems necessary to separately
  postulate a low entropy initial state. 
\end{abstract}

\maketitle

The comment by Jennings and Rudolph~\cite{comment} provides a clever
example where (classical) mutual information (CMI) may increase even
in situations where quantum mutual information (QMI) decreases. This
is not unexpected as the QMI is only an upper bound to the CMI, as is
well known and also stated in my paper~\cite{entropy}. However, their
analysis does point out a flaw in my argument: I had implicitly
assumed that a decrease of QMI between two systems always derives from
the nullification of the QMI among two of the degrees of freedom in
the two systems, which would indeed entail a decrease of CMI. I
acknowledge that this hidden assumption is unwarranted.

The main claims contained in my paper (summarized in its abstract)
are, however, not affected by this: the argument still goes through,
although it is now clear that CMI is not the appropriate measure of
information here. This is not too surprising, as in the quantum
framework that is employed in my paper it is the QMI that measures the
total amount of correlations (quantum {\em and}
classical)~\cite{popescuw}. Namely, the quantitative analysis
in~\cite{entropy} should end with Eq.~(2): a decrease of entropy of a
system that does not follow from dumping entropy to an environment
must entail a decrease of the {\em quantum} mutual information between
the system and its observer.

It might be useful to reiterate the argument of~\cite{entropy}. Any
information an observer has on a system must be encoded into some
correlation between the system and a degree of freedom related to the
observer. If one considers also the purification space (we assume that
the universe is in a pure state, so such purification always exists),
then this correlation consists of entanglement between the systems
involved: the only correlations in a pure state arise from
entanglement. If one of the systems is an observer (all observers
agree with each other when they witness the same
events~\cite{everett}), then this entanglement entails from the
observer's point of view a probability stemming from the Born rule,
and thus an associated entropy. The reduction of this entropy is in
one-to-one correspondence with the reduction of the entanglement (as
long as entropy is not transferred to the environment). This then
entails a reduction of the information~\footnote{Note that, as
  Jennings and Rudolph's example clearly shows, an interaction between
  system and environment can change the QMI between system and
  observer without affecting the observer's entropy or state.}.

Only in quantum mechanics, correlation in a pure state is due to
entanglement which in turn is responsible for the entropy of a
subsystem of such a system. Thus correlation and entropy are in
one-to-one correspondence. This is not true in classical mechanics,
where correlations can be increased at will without affecting the
entropy. The fact that the variation in mutual information (a measure
of correlation~\cite{popescuw}) is equal to the local variations of von
Neumann entropy (which can be put in one-to-one correspondence with
thermodynamic entropy through Szilard engines or Maxwell demons),
whenever the environment entropy is unaffected, is not an empty
statement, contrary to Jennings and Rudolph's claims: for example, an
analogous statement is not true in classical mechanics.
          
In the decoherence language~\cite{zeh}, my argument entails that, in
order for a recoherence (namely a fusion of different Everett
branches~\cite{everett}) to happen from the point of view of the
observer, all entanglement that built up\comment{ among him/her/it and
  the other systems} during the decoherence event must be eliminated.
This decorrelates the subsystems so that they are oblivious of having
previously interacted during the decoherence process: no information
on that interaction must be left (as it would prevent recoherence).
Namely, any recoherence cannot leave trace of the previous
decoherence.

One last important point which was unfortunately not addressed in my
paper~\cite{entropy} must be emphasized. A good solution to the
arrow-of-time dilemma must explain two main features: why entropy only
increases and why it was so low in the initial state of the universe.
My argument provides an answer to both: entropy only increases because
any decrease cannot leave information of it having happened, and thus
looking further into the past all observers will subjectively (but
collectively) see entropy steadily decreasing for decreasing time up
to a low initial entropy.  However, my argumentfails to describe the
present. In fact, without any prior assumption on the initial entropy,
one must assume that the universe is in a highly-probable high-entropy
state~\cite{popescu} at every time (though it subjectively does not
appear so when looking to the past).  Then, any observer should see a
high entropy state in his/her/its present. This is not what we
observe: the current state of the universe has very high entropy, but
is far from thermal equilibrium.

This implies that my argument is not a complete solution for the
arrow-of-time dilemma: it seems that a low entropy initial state (more
precisely, a pure state where most subsystems were factorized and
highly symmetric~\cite{sethzur}) must be separately postulated to
account for phenomena we see. My argument does, however, give a
description of the second law from a quantum observer's point of view,
which subjectively sees entropy constantly increasing even in extreme
situations of a near-thermal equilibrium universe.

\comment{In usual everyday situations, recoherence events that involve an
observer are extremely rare (or, rather, practically impossible). This
implies that, although correct in principle, my argument is
inconsequential in practice: most entropy-increasing events can be
explained solely from the fact that for macroscopic systems
decoherence is much more probable than recoherence.  This asymmetry
can be seen as stemming from asymmetric boundary conditions {\em \`a
  la} Boltzmann (more precisely, from the fact that in the initial
state of the universe most subsystems were factorized and highly
symmetric~\cite{sethzur}). The mechanism I describe in my paper
becomes relevant only when recoherence events are comparable in
probability with decoherence events, namely in a pure-state universe
near a heat-death state where subsystems are almost maximally
correlated.}

\comment{This implies that my argument cannot be seen as a complete solution
for the arrow-of-time dilemma: even though it might explain how
entropy-increasing events are singled out from any observer's point of
view, it cannot account for the humongous difference between initial
and current entropy of the universe.}

I thank Prof.~H.~D.~Zeh for his patient and useful feedback and S.
Lloyd for useful discussions.

\end{document}